\documentclass[aps,floatfix,prd,showpacs]{revtex4}
\usepackage{graphicx}
\usepackage{dcolumn}
\usepackage{bm}

\usepackage{amsmath,amsthm,amssymb,epsfig,alltt}

\begin{document}

\def\a{\alpha}
\def\b{\beta}
\def\d{{\delta}}
\def\l{\lambda}
\def\e{\epsilon}
\def\p{\partial}
\def\m{\mu}
\def\n{\nu}
\def\t{\tau}
\def\th{\theta}
\def\s{\sigma}
\def\g{\gamma}
\def\G{\Gamma}
\def\o{\omega}
\def\r{\rho}
\def\z{\zeta}
\def\D{\Delta}
\def\half{\frac{1}{2}}
\def\hatt{{\hat t}}
\def\hatx{{\hat x}}
\def\hatp{{\hat p}}
\def\hatX{{\hat X}}
\def\hatY{{\hat Y}}
\def\hatP{{\hat P}}
\def\haty{{\hat y}}
\def\whatX{{\widehat{X}}}
\def\whata{{\widehat{\alpha}}}
\def\whatb{{\widehat{\beta}}}
\def\whatV{{\widehat{V}}}
\def\hatth{{\hat \theta}}
\def\hatta{{\hat \tau}}
\def\hatrh{{\hat \rho}}
\def\hatva{{\hat \varphi}}
\def\barx{{\bar x}}
\def\bary{{\bar y}}
\def\barz{{\bar z}}
\def\baro{{\bar \omega}}
\def\barpsi{{\bar \psi}}
\def\sp{\sigma^\prime}
\def\nn{\nonumber}
\def\cb{{\cal B}}
\def\2pap{2\pi\alpha^\prime}
\def\wideA{\widehat{A}}
\def\wideF{\widehat{F}}
\def\beq{\begin{eqnarray}}
 \def\eeq{\end{eqnarray}}
 \def\4pap{4\pi\a^\prime}
 \def\xp{{x^\prime}}
 \def\sp{{\s^\prime}}
 \def\ap{{\a^\prime}}
 \def\tp{{\t^\prime}}
 \def\zp{{z^\prime}}
 \def\xpp{x^{\prime\prime}}
 \def\xppp{x^{\prime\prime\prime}}
 \def\barxp{{\bar x}^\prime}
 \def\barxpp{{\bar x}^{\prime\prime}}
 \def\barxppp{{\bar x}^{\prime\prime\prime}}
 \def\barchi{{\bar \chi}}
 \def\baro{{\bar \omega}}
 \def\bpsi{{\bar \psi}}
 \def\barg{{\bar g}}
 \def\barz{{\bar z}}
 \def\bareta{{\bar \eta}}
 \def\ta{{\tilde \a}}
 \def\tb{{\tilde \b}}
 \def\tc{{\tilde c}}
 \def\tz{{\tilde z}}
 \def\tJ{{\tilde J}}
 \def\tpsi{\tilde{\psi}}
 \def\tal{{\tilde \alpha}}
 \def\tbe{{\tilde \beta}}
 \def\tga{{\tilde \gamma}}
 \def\tchi{{\tilde{\chi}}}
 \def\barth{{\bar \theta}}
 \def\bareta{{\bar \eta}}
 \def\barom{{\bar \omega}}
 \def\bole{{\boldsymbol \epsilon}}
 \def\bolth{{\boldsymbol \theta}}
 \def\bomega{{\boldsymbol \omega}}
 \def\bolmu{{\boldsymbol \mu}}
 \def\bola{{\boldsymbol \alpha}}
 \def\bolb{{\boldsymbol \beta}}
 \def\bolX{{\boldsymbol X}}
 \def\mathN{{\boldsymbol n}}
 \def\bba{{\boldsymbol a}}
 \def\bby{{\boldsymbol y}}
 \def\bbp{{\boldsymbol p}}
 \def\bbphi{{\boldsymbol \phi}}
 \def\bbA{{\boldsymbol A}}
 \def\mathP{{\mathbb P}}
 \def\mathN{{\boldsymbol N}}
 \def\mathN{{\mathbb N}}
 \def\bbP{{\boldsymbol P}}


\title{Saving Tachyons in Open String Theory}
\author{Taejin Lee}
\affiliation{
Department of Physics, Kangwon National University, Chuncheon 24341
Korea}

\email{taejin@kangwon.ac.kr}


\begin{abstract}
Using string scattering amplitudes of open bosonic string, we construct a local field theoretical action 
for tachyon fields and look for a possibility of homogeneous static vacuum which may result from tachyon condensation. 
Cubic local interactions between various particles, belonging to the particle spectrum of string may be directly followed from three-string scattering amplitude. These cubic local interactions may generate perturbative non-local four-particle interactions, which may contribute to four-string scattering amplitude. 
The Veneziano amplitude, describing four-tachyon scattering, expanded in terms of $s$-channel poles was compared
with the four-tachyon scattering amplitudes in $s$-channel generated perturbatively and it was found
that a quartic potential term is needed in the local field theoretical action, which describes open string theory effectively in the 
low energy regime. With this quartic term, the tachyon potential has a stable minimum point and the tachyon field may condensate. As a result, 
both tachyon and gauge fields become massive at Planck scale and completely disappear from the low energy  particle spectrum.
\end{abstract}



\maketitle

\section{Introduction}

The birth of string theory is widely credited to the Veneziano amplitude \cite{Veneziano68}, which was later known to describe 
four-tachyon-scattering of open string theory \cite{DiVecchia2008}. The four-point amplitude of scalar particles possesses the 
crossing symmetry and exhibits Regge behavior at high energy, which are important in understanding the strongly interacting processes. 
Shortly after the advent of Veneziano amplitude, Virasoro \cite{Virasoro69} and Shapiro \cite{Shapiro69} constructed another crossing 
symmetric amplitude which was named as the Shapiro-Virasoro amplitude. It later turned out that the underlying structures behind these 
amplitudes were quantum relativistic strings and the Veneziano amplitude and the Shapiro-Virasoro amplitude correspond to 
four-tachyon-scattering amplitudes of open string and closed string respectively. Since the presence of tachyons indicates that the system is unstable, 
bosonic open string theory has not been considered as a realistic model. In an effort to construct a more 
stable realistic model, super-symmetric string theories have been proposed by Neveu,  Schwarz \cite{NeveuNP1971,
NeveuPR1971} and Ramond \cite{Ramond1971}. In the case of super-symmetric theory, known as the NSR (Neveu-Schwarz-Ramond) model, one can consistently remove the unwanted sector, containing tachyons by applying the GSO
( Gliozzi-Scherk-Olive) projection \cite{GSO1977}.
However, the GSO projection must be imposed by hand to obtain a stable vacuum: This may be considered as a 
drawback of this procedure. It is more desirable to find a dynamical process to remove the unstable sector. 

More recent studies on tachyons in string theory, which utilized tachyons to describe unstable $D$-brane systems
have focused on time-dependent or inhomegeneous tachyon condensation: 
Unstable $D$-branes may decay into stable lower dimensional $D$-brane systems \cite{sen1998}.
As tachyon condensates on a $D$-brane, an exactly marginal boundary term may be induced on the world-sheet string action. 
The conformal field theory with this marginal boundary term has been studied extensively by using the boundary string field theory \cite{SenZ2000,GerasimovS2000,KutasovM2000,KrausL2001,TakayanagiU2001} and the boundary state formulation 
\cite{Tlee:01nc,Tlee:01os,Tlee:01one,TLeeSem2005,Hasselfield2006,TLee2006JHEP,TLee2008JHEP,TLee2009}, 
because it may describe time evolution of unstable $D$-brane systems \cite{Sen:2002nu}. 

This present work studied scattering amplitudes of open string theory in the proper-time gauge 
\cite{TLee88ann,Lee2016i,Lee2017d,TLee2017cov,Lai2017S,TLeeEPJ2018,TLeeEnt2018,TLeeGauge2018}, 
to construct the local action for tachyon fields. 
Using the resultant local action for tahcyon, we look for a possibility of static homogeneous tachyon condensation
which may result in a homegeneous static vacuum. It was Bardakci \cite{Bardakci1974} who first
studied the spontaneous symmetry breaking due to homogeneous static tachyon condensation in string theory,
which used to be called dual model at that time. If the tachyon condensates homogeneoously to 
yield a stable static vacuum around which tachyon and massless particles become massive at Planck scale,
we may not need to impose the GSO projection to remove the unstable sector. Because the particles 
in the uwanted sector would disappear completely from the low energy particle spectrum.
Here, we look for such a possibility, constructing tachyon potential, which is compatible with string scattering 
amplitudes. 
  
The three-string scattering amplitudes completely fix cubic
couplings of various particles which appear in the spectrum of free open string. Then, using the 
usual Feynman diagrams, the four-particle scattering amplitudes, generated perturbatively 
by the cubic interactions were evaluated and compare to the four-string scattering amplitudes. There are 
some differences between the four-particle scattering amplitudes generated by cubic interactions and
corresponding ones obtained from string scattering amplitudes. This discrepancy may be due to 
contributions of local four-particle interactions to four-particle scattering amplitudes, 
included in the string scattering amplitudes. Therefore, it is possible to completely fix the 
local four-particle interactions, which may arise in string theory. By applying this procedure to the 
tachyon scattering amplitudes, it is possible to find a local quartic tachyon potential that may lead the tachyon field
to condensate. The tachyon condensation may stabilize the $D$-brane system to turn tachyons into massive particles.
After condensation occurs, gauge fields also acquire masses and the particle spectrum of open string contains massive particles only.

\section{Cubic Interaction Terms from Three-String-Scattering Amplitude}

\begin{figure}[htbp]
\begin {center}
\epsfxsize=0.8\hsize

\epsfbox{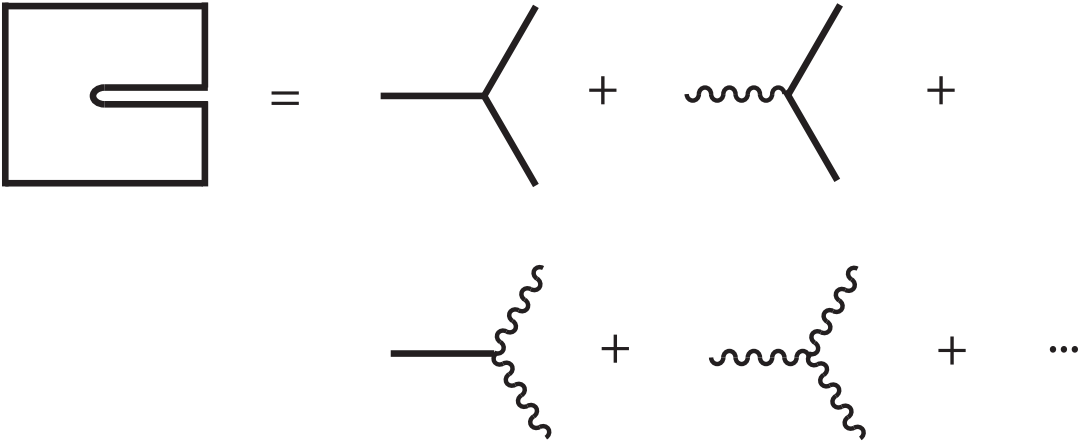}
\end {center}
\caption {\label{threeopen} Three-string scattering and cubic couplings.}
\end{figure}

The cubic interaction terms of particles may be directly read off from three-string-scattering amplitude. 
The three-string-scattering amplitude may also be obtained by evaluating the Polyakov string path integral \cite{Polyakov1981} on 
a Riemann surface which depicts a string world sheet of three-open-string scattering. As discussed in previous
works \cite{Lee2016i,TLee2017cov}, it is convenient to choose a proper-time gauge to cast string scattering amplitude into 
the Feynman-Schwinger proper-time representation of quantum field theory. 
The three-string scattering amplitude may be written in the proper-time gauge \cite{Lee2017d,TLee2017cov} as 
\begin{subequations}
\beq \label{3string}
{\cal I}_{[3]} &=& \int \prod_{r=1}^3 dp^{(r)} \d \left(\sum_{r=1}^3 p^{(r)} \right)  \frac{2g}{3} \langle \Psi_1, \Psi_2, \Psi_3 \vert E[1,2,3] \vert 0 \rangle, \\
E[1,2,3] \vert 0 \rangle
&=& \exp \,\Biggl\{ \frac{1}{2} \sum_{r,s =1}^3 \sum_{n, m \ge 1} \bar N^{rs}_{nm} \, 
\a^{(r)\dagger}_{n\m} \a^{(s)\dagger}_{m\n} \eta^{\m\n} + 
\sum_{r=1}^3 \sum_{n \ge 1} \bar N^r_n \a^{(r)\dag}_{n\m} \bbP^\m \nn\\
&& +\t_0 \sum_{r=1}^3 \frac{1}{\a_r} \left(\frac{(p^{(r)}_\m p^{(r)\m}}{2} -1 \right)
\Biggr\} \vert 0 \rangle, \label{E3}
\eeq
\end{subequations}
where $\bbP= p^{(2)}- p^{(1)} $.
Here $\bar N^{rs}_{nm}$ and $\bar N^r_n$ are the Neumann functions of three-string scattering in the proper-time gauge. 
The scattering amplitudes of three-particle-scattering on a space-filling $D$-brane, which can be decomposed into three-particle scattering amplitudes,
can be obtained by choosing the external string state as follows
\beq \label{external3}
\langle \Psi^{(1)}, \Psi^{(2)}, \Psi^{(3)} \vert 
= \langle 0 \vert \prod_{r=1}^3 \left(\bbphi(r) + \bbA(r)\right)
\eeq 
where $\bbphi(r) = \phi(p^{(r)}), ~\bbA(r) = A_\m(p^{(r)}) a^{(r)\m}_{1}, ~ r= 1, 2, 3$.
The cubic couplings of gauge fields $A^\m$ and tachyon field, $\phi$ are directly deduced from 
the three-particle scattering amplitudes. Fig. \ref{threeopen} illustrates an expansion of three-string scattering amplitude
in terms of three-particle scattering amplitudes. 

The three-tachyon scattering amplitude readily follows from Eq. (\ref{3string}) and Eq. (\ref{external3}) :
\beq \label{s3low}
{\cal I}_{\phi\phi\phi}
&=& \frac{2g}{3} \int \prod_{r=1}^3 dp^{(r)} \d \left(\sum_{r=1}^3 p^{(r)} \right) \,\prod_{r=1}^3\Bigl(\phi(p^{(r)})\Bigr ) .
\eeq
It can be also understood as a local cubic term in the field theoretical action for tachyon
\beq
{\cal A}_{\phi\phi\phi} = \frac{2g}{3} \int d^d x\, \phi^3 . 
\eeq 
The scattering amplitude for two-tachyon and one-gauge particle is given by
\beq
{\cal I}_{\phi\phi A} &=& \frac{2g}{3}\int \prod_{r=1}^3 dp^{(r)} \d \left(\sum_{r=1}^3 p^{(r)}\right) \, 
\langle 0 \vert \bigl(\bbphi(1)\bbphi(2) \bbA(3) + \bbphi(1)\bbA(2) \bbphi(3)+ \bbA(1)\bbphi(2)\bbphi(3) 
\bigr) \nn\\
&& e^{\t_0\sum_{r=1}^3 \frac{1}{\a_r} \left(\frac{(p^{(r)})^2}{2}-1 \right)} \left(\sum_{r=1}^3\bar N^r_1 a^{(r)\dag}_{1} \cdot \bbP\right) \vert 0 \rangle ,
\eeq
where $\t_0 = -2 \ln 2$ and $\bar N^1_1 = \bar N^2_1 = \frac{1}{4}, \bar N^3_1 = -1$ \cite{Lee2016i}. 
Imposing the on-shell conditions for tachyon and gauge fields leads to:
\beq
{\cal I}_{\phi\phi A} 
&=& g \int \prod_{r=1}^3 dp^{(r)} \d \left(\sum_{r=1}^3 p^{(r)}\right) \, 
\left(p^{(1)}_\m -p^{(2)}_\m\right)\phi(1) \phi(2) A(3)^\m .
\eeq 
Note that if $\phi$ is real, this term may vanish. For open string on $N$ multiple $D$-branes, the tachyon and gauge fields may be represented by $U(N)$ matrix valued fields. Then the scattering amplitude, ${\cal I}_{\phi\phi A}$ is not vanishing and 
may be written as 
\beq
{\cal I}_{\phi\phi A} &=& g \int \prod_{r=1}^3 dp^{(r)} \d \left(\sum_{r=1}^3 p^{(r)}\right) \, 
p^{(1)}_\m {\rm tr} \left(\phi(1) \left[\phi(2), A(3)^\m \right] \right).  
\eeq

Similarly, scattering amplitude of one-tachyon and two-gauge particle, ${\cal I}_{\phi A A}$
may be evaluated 
\beq\label{Ipaa}
{\cal I}_{\phi A A} &=& \frac{2g}{3}\int \prod_{r=1}^3 dp^{(r)} \d \left(\sum_{r=1}^3 p^{(r)}\right) \, 
\, \langle 0 \vert  \bigl(2\bbphi(1)\bbA(2) \bbA(3) + 2\bbA(1)\bbphi(2) \bbA(3)+ 2^4\bbA(1)\bbA(2)\bbphi(3) 
\bigr)\nn\\
&&\left\{\half \sum_{r,s=1}^3 \bar N^{rs}_{11} a^{(r)\dagger}_{1} \cdot a^{(s)\dagger}_{1} + 
\frac{1}{2!}\left(\sum_{r=1}^3\bar N^r_1 a^{(r)\dagger}_{1} \cdot \bbP \right)^2
\right\} \vert 0 \rangle \nn\\
&=& g \int d^d x \, \left(\phi A_\m A^\m + \phi\, \p_\m A^\n \p_\n A^\m \right).
\eeq
The second term may be considered as an $\ap$-correction to the cubic coupling of $\phi\phi A$. 
As for cubic coupling of gauge fields, recalling that the scattering amplitude of three-gauge particle has been
evaluated before in Refs. \cite{Lee2016i,Lee2017d,TLee2017cov}.
The three-gauge particle scattering amplitude was found to be
\begin{subequations}
\beq
{\cal I}_{AAA} &=& {\cal I}_{AAA}^{(0)} + {\cal I}_{AAA}^{(1)}, \\
{\cal I}_{AAA}^{(0)} &=& \frac{2^4g}{3}\int \prod_{r=1}^3 dp^{(r)} \d \left(\sum_{r=1}^3 p^{(r)}\right) \, 
\, \langle 0 \vert \bbA(1)\bbA(2) \bbA(3) \left\{\half \sum_{r,s=1}^3 \bar N^{rs}_{11} a^{(r)\dagger}_{1} \cdot a^{(s)\dagger}_{1}
\left(\sum_{r=1}^3\bar N^r_1 a^{(r)\dagger}_{1} \cdot \bbP \right)\right\}\vert 0 \rangle . \\
{\cal I}_{AAA}^{(1)} &=& \frac{2^4g}{3}\int \prod_{r=1}^3 dp^{(r)} \d \left(\sum_{r=1}^3 p^{(r)}\right) \, \langle 0 \vert \bbA(1)\bbA(2) \bbA(3) \left\{
\frac{1}{3!}\left(\sum_{r=1}^3\bar N^r_1 a^{(r)\dagger}_{1} \cdot \bbP \right)^3
\right\} \vert 0 \rangle \nn\\
&=& \frac{g}{3}i \int d^d x \, F_\m{}^\n F_\n{}^\l F_\l{}^\m.
\eeq 
\end{subequations}
The first cubic interaction term of gauge field, ${\cal I}_{AAA}^{(0)}$ vanishes for gauge field ($N=1$) on a single $D$-brane. 
This is consistent with absence of cubic term in the action of $U(1)$ gauge field. For gauge fields on $N$ multiple $D$-branes, 
which represented by $U(N)$ non-Abelian fields, this term does not vanish and 
may be identified as the cubic interaction part of the covariant Yang-Mills action
\beq
{\cal I}_{AAA}^{(0)} 
&=& g i \int d^d x \,\text{tr} \left(\p_\m A_\n - \p_\n A_\m \right) \left[ A^\m, A^\n \right] \label{gauge3}.
\eeq
The second term, ${\cal I}_{AAA}^{(1)}$ is in consistence with the $\ap$-correction to the three-gauge field interaction term, 
which was obtained by previous approaches \cite{Neveu72,Scherk74,Tseytlin86,Coletti03}.

\section{Veneziano Amplitude and Feynman Diagrams}

\begin{figure}[htbp]
\begin {center}
\epsfxsize=0.8\hsize

\epsfbox{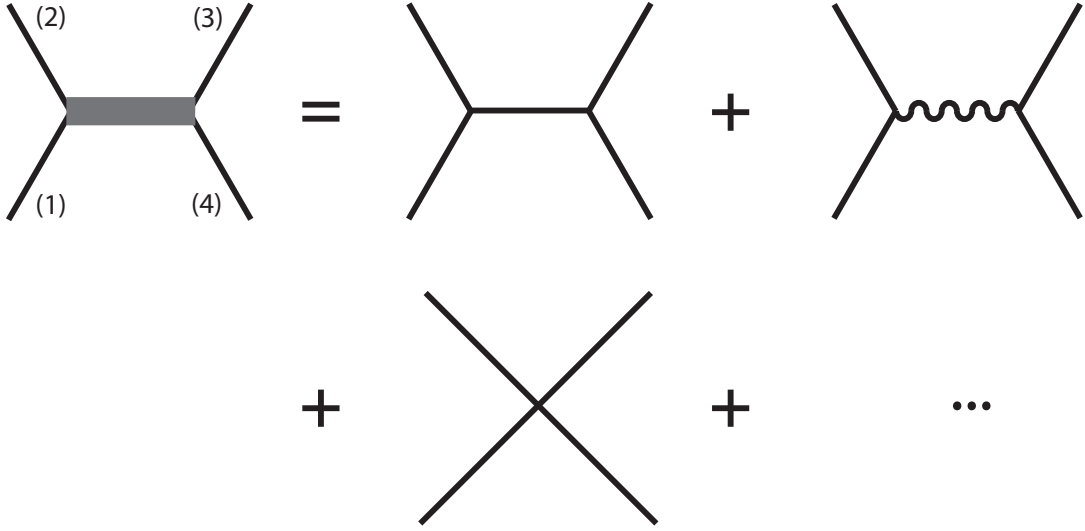}
\end {center}
\caption {\label{star2} Feynman diagrams of four-tachyon scattering amplitude.}
\end{figure}

From the string theory, the four-tachyon-scattering amplitude is described by the 
Veneziano amplitude:  
The four-tachyon scattering amplitude in $(s,t)$ channel may be written in terms of Mandelstam variables as 
\beq
{\cal F}_{[4]}^{\rm tachyon} (s,t) 
=  g^2 \frac{\G\left(-\frac{s}{2} -1 \right) 
\G\left(-\frac{t}{2} -1 \right)}{\G\left(-\frac{s}{2} - \frac{t}{2} -2 \right)} 
\eeq
where $ s+t+u = -8$.
It may be expanded with Regge poles in $s$-channel, 
\beq
{\cal F}_{[4]}^{\rm tachyon} (s,t) 
&=& - g^2 \sum_{n=0}^\infty \frac{\left(\a(t) + 1 \right) \left(\a(t) + 2 \right) 
\cdots \left(\a(t) + n \right)}{n!} \frac{1}{\a(s) - n} \nn\\
&=&  - \frac{2g^2 }{s+2} - \frac{g^2\left(t+4\right) }{s} - \frac{g^2}{4} \frac{(t+4)(t+6)}{s-2}+ \cdots 
\eeq 
where $\a(s) = 1 + s/2$ and $\a(t) = 1 + t/2$.
To be precise, the four-tachyon scattering amplitude of string theory, deduced from the Veneziano 
amplitude should be written as 
\beq \label{schannel}
{\cal I}_{\phi^4}^{\rm Veneziano} &=& \int \prod_{r=1}^4 dp^{(r)}\, \d\left(\sum_{r=1}^4 p^{(r)} \right)\, \left(\phi(p^{(1)}) \phi(p^{(2)}) \phi(p^{(3)})\phi (p^{(4)})\right){\cal F}_{[4]}^{\rm tachyon} (s,t) \nn\\
&=& \int \prod_{r=1}^4 dp^{(r)}\, \d\left(\sum_{r=1}^4 p^{(r)} \right)\, \left(\phi(p^{(1)}) \phi(p^{(2)}) \phi(p^{(3)})\phi (p^{(4)})\right) \left\{- \frac{2g^2 }{s+2} - \frac{g^2\left(t+4\right) }{s} + \cdots \right\}.
\eeq  
This expansion of the Veneziano amplitude shall be compared with the four-tachyon-scattering amplitudes
generated perturbatively by cubic interaction terms obtained in the previous section.
Fig.\ref{star2} pictorially depicts expansion of the Veneziano amplitude in terms of Regge poles in $s$-channel. 
The Veneziano amplitude contains the scattering amplitude due to a local quartic tachyon interaction in addition 
to those generated by cubic particle interactions perturbatively. 

The first term in the expansion Eq.(\ref{schannel}) is in perfect agreement with the perturbative 
scattering amplitude of four tachyons mediated by tachyon propagator as depicted in Fig. \ref{star2}. 
\beq
\left . {\cal I}^{\rm pert}_{\phi^4} \right\vert_{\rm tachyon} &=& 2 g^2 \int \prod_{r=1}^4 dp^{(r)}\, \d\left(\sum_{r=1}^4 p^{(r)} \right)\, \left(\phi(p^{(1)}) \phi(p^{(2)}) \frac{1}{\left(p^{(1)}+p^{(2)}\right)^2-2} \phi(p^{(3)})\phi (p^{(4)})\right) \nn\\
&=& \int \prod_{r=1}^4 dp^{(r)}\, \d\left(\sum_{r=1}^4 p^{(r)} \right)\, \left(- \frac{2g^2}{s+2}\right)\, \left(\phi(p^{(1)}) \phi(p^{(2)}) \phi(p^{(3)})\phi (p^{(4)})\right) .
\eeq 
The second term in the expansion may be reproduced by a perturbative the Feynman diagram of four-tachyon 
scattering amplitude mediated by a massless gauge particle. However, as observed in the last section, 
if the tachyon field $\phi$ is a real scalar, the cubic coupling between two tachyons and one gauge field vanishes so that it is impossible to 
generate perturbatively the four-tachyon-scattering amplitude mediated by a massless gauge particle. 
This difficulty may be resolved if we represent the tachyon field by a complex field. Then, the scattering amplitude, ${\cal I}_{\phi\phi A}$,
rewritten in terms of the complex tachyon field, is not vanishing  
\beq
{\cal I}_{\phi\phi A} &=& g \int \prod_{r=1}^3 dp^{(r)} \d \left(\sum_{r=1}^3 p^{(r)}\right) \,  \left\{\left(p^{(1)}_\m -p^{(2)}_\m\right)\phi^\dag(1) \phi(2) A(3)^\m
\right\} \nn\\
&=& -i \int d^d x \, \left(\p_\m \phi^\dag \phi - \phi^\dag \p_\m \phi \right) A^\m .
\eeq
This cubic coupling between tachyon and gauge fields generates perturbatively the four-tachyon scattering amplitude written as: 
\beq
\left .{\cal I}_{\phi^4}^{\text{pert}} \right\vert_{\rm gauge} &=& - \frac{g^2}{2} \int \prod_{r=1}^4 dp^{(r)}\, \d\left(\sum_{r=1}^4 p^{(r)} \right)\, \left(\phi(p^{(1)})^\dag \phi(p^{(2)}) \phi(p^{(3)})\phi^\dag (p^{(4)})\right) \nn\\
&& (p^{(1)}-p^{(2)})_\m\frac{1}{\left(p^{(1)}+p^{(2)}\right)^2} 
\left(\eta^{\m\n} - \frac{\left(p^{(1)}+p^{(2)}\right)^\m \left(p^{(1)}+p^{(2)}\right)^\n}{\left(p^{(1)}+p^{(2)}\right)^2} \right) 
(p^{(3)}-p^{(4)})_\n \nn\\
&=& - \frac{g^2}{2} \int \prod_{r=1}^4 dp^{(r)}\, \d\left(\sum_{r=1}^4 p^{(r)} \right)\, \left(\phi(p^{(1)})^\dag \phi(p^{(2)}) \phi(p^{(3)})\phi^\dag (p^{(4)})\right)
\left\{ \frac{2(t+4) +s}{s} \right\}
\eeq 
where the gauge field propagator is chosen by 
$G_{\m\n}(p) = - \frac{1}{p^2} \left(\eta^{\m\n} - \frac{p^\m p^\n}{p^2} \right)$. 
The Mandelstam variables are defined for four-tachyon-scattering by 
\begin{subequations}
\beq
s &=& - \left(p^{(1)}+p^{(2)}\right)^2 = - \left(p^{(3)}+p^{(4)}\right)^2 
= -2 p^{(1)} \cdot p^{(2)} -4 =-2 p^{(3)} \cdot p^{(4)} -4 ,\\
u &=& - \left(p^{(1)}+p^{(3)}\right)^2 = - \left(p^{(2)}+p^{(4)}\right)^2
= -2 p^{(1)} \cdot p^{(3)} -4 =-2 p^{(2)} \cdot p^{(4)} -4, \\
t &=& - \left(p^{(1)}+p^{(4)}\right)^2 = - \left(p^{(2)}+p^{(3)}\right)^2
= -2 p^{(1)} \cdot p^{(4)} -4 =-2 p^{(2)} \cdot p^{(3)} -4 
\eeq 
\end{subequations}
$s+t+u =-8$.
Comparing the string scattering amplitude of four-tachyon scattering with a massless pole, 
deduced from the Veneziano amplitude with that of four-tachyon mediated by a massless gauge field, 
we may identify the local four-tachyon interaction:
\beq
{\cal I}^{\rm local}_{\phi^4} &=&\left . {\cal I }^{\text{string}}_{\phi^4}\right\vert_{\rm gauge}- \left .{\cal I}^{\text{pert}}_{\phi^4} \right\vert_{\rm gauge}\nn\\
&=& - g^2\int \prod_{r=1}^4 dp^{(r)}\, \d\left(\sum_{r=1}^4 p^{(r)} \right)\, \frac{(t+4)}{s}
\, \left(\phi(p^{(1)})^\dag \phi(p^{(2)}) \phi^\dag(p^{(3)})\phi(p^{(4)})\right) \nn\\
&&+ 
\frac{g^2}{2}\int \prod_{r=1}^4 dp^{(r)}\, \d\left(\sum_{r=1}^4 p^{(r)} \right)\, \left(\frac{2(t+4) + s}{s} \right)\, \left(\phi(p^{(1)})^\dag \phi(p^{(2)}) \phi^\dag(p^{(3)})\phi(p^{(4)})\right) \nn\\
&=& \frac{g^2}{2} \int \prod_{r=1}^4 dp^{(r)}\, \d\left(\sum_{r=1}^4 p^{(r)} \right)\, \left(\phi(p^{(1)})^\dag \phi(p^{(2)}) \phi^\dag(p^{(3)})\phi(p^{(4)})\right) \nn\\
&=& \frac{g^2}{2} \int d^d x \, \left(\phi^\dag \phi\right)^2 .  
\eeq  
It was found that a local four-tachyon interaction term is needed to match the four-tachyon scattering amplitude of string theory with a local perturbative field theory. It is noteworthy that this quartic term of tachyon potential has been also found in D-brane-anti D-brane systems \cite{Hatefi2013,Hatefi2016,Hatefi2017,Hatefi2019}. 
Accordingly, the cubic interaction ${\cal I}_{\phi A A}$, Eq.(\ref{Ipaa}) may be rewritten in terms of the complex tachyon field as follows 
\beq 
{\cal I}_{\phi A A} &=& \frac{g}{2} \int d^d x \, \left(\left(\phi + \phi^\dag \right) A_\m A^\m + \left(\phi+ \phi^\dag\right)\, \p_\m A^\n \p_\n A^\m \right).
\eeq  
Likewise, the tachyon potential, including the cubic and quartic terms, may be given by 
\beq
V_T(\phi) &=& - \phi^\dag \phi - \frac{g}{3} \left(\phi^\dag + \phi \right) \left(\phi^\dag \phi \right)+ \frac{g^2}{2} \left(\phi^\dag \phi\right)^2.
\eeq

\section{Tachyon Condensation and Homogeneous Static Vacuum}

\begin{figure}[htbp]
\begin {center}
\epsfxsize=0.5\hsize

\epsfbox{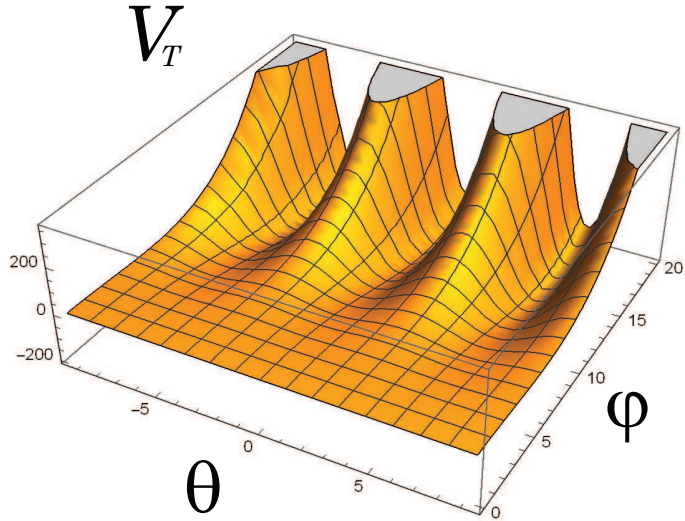}
\end {center}
\caption {\label{tachpotential} Tachyon potential in a covering space.}
\end{figure}

Having fixed the tachyon potential, we carry out functional analysis to find stable minima. 
It may be convenient to introduce two real fields, $\th$ and $\varphi$ to express the complex tachyon field,
$\phi (x) = e^{i\th(x)} \varphi(x)$, to analyze the tachyon potential. 
In terms of two real fields, the tachyon potential may be written as follows:
\beq
V_T(\th, \varphi) = - \varphi^2 - \frac{2g}{3} \cos \th\, \varphi^3 + \frac{g^2}{2} \varphi^4 . 
\eeq 
Extrema of the potential are determined by 
\begin{subequations}
\beq
\frac{\p V_T}{\p \varphi} &=& -2 \varphi -2g \cos \th \varphi^2 + 2g^2 \varphi^3 =0, \\
\frac{\p V_T}{\p \th} &=& \frac{2g}{3} \sin \th\, \varphi^3 = 0. 
\eeq 
Fig. \ref{tachpotential} depicts the tachyon potential in a covering space, showing that 
the tachyon potential has three extrema with $\th = 2n \pi$, $n \in Z$ where
\beq
\varphi= \frac{\left(1-\sqrt{5}\right)}{2g}, ~~\varphi = 0, ~~\varphi= \frac{\left(1+\sqrt{5}\right)}{2g}.
\eeq 

\end{subequations}

\begin{figure}[htbp]
\begin {center}
\epsfxsize=0.5\hsize

\epsfbox{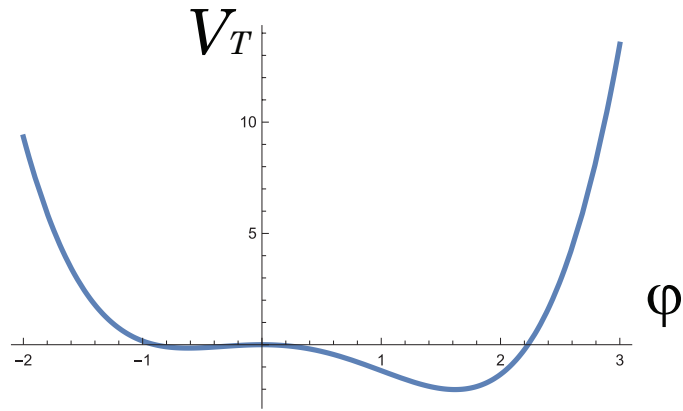}
\end {center}
\caption {\label{tachyon1a} Tachyon potential along the line, $\th=0$.}
\end{figure}

A stable minimum is located at $\varphi= \varphi_0 = \frac{\left(1+\sqrt{5}\right)}{2g}$, $\th =0$. 
The tachyon potential along the line with $\th =0$ is portrayed in Fig. \ref{tachyon1a}.
It is clear from the pictures that  tachyon field may condensate; $ \left<\phi \right> = \varphi_0$. 
Near the stable minimum point, if the tachyon field $\phi$ is expanded, 
\beq
\phi = e^{i\th} \left(\varphi_0 + \varphi \right), 
\eeq 
the Lagrangian for tachyon, including the potential $V_T$, may be written by
\beq \label{tachyonmass}
L &=& - \frac{1}{2} \bigl(\p \varphi \p \varphi + (\varphi_0)^2 \p \th \p \th \bigr) -
\left(\frac{2+ \sqrt{5}}{3g^2}\right) \th^2 - \frac{(5+ \sqrt{5})}{2} \varphi^2
- V_T(0, \varphi_0).
\eeq 
It follows from Eq.(\ref{tachyonmass}) that the tachyon fields represented by $\th$ and $\varphi$ become
very massive and completely disappear from the physical low energy spectrum of particles:
Scaling $\th \rightarrow \th/\varphi_0$, we find 
$m_\th^2 = \frac{1+\sqrt{5}}{6}, ~~ m_\varphi^2 = 5+ \sqrt{5}$.
It is also notable that the gauge particle, $A_\m$ becomes a massive one at Planck scale
with $m^2_{\rm A} = (1+ \sqrt{5})/2$ as a result of the cubic interaction, $\phi A_\m A^\m$. 
(After the tachyon condensation, a quartic interaction of type $\vert\phi\vert^2 A_\m A^\m$ additionally contributes 
to the mass of gauge particle.)
Thus, due to the tachyon condensation, both scalar tachyons and gauge particles acquire 
masses at Planck scale and as a result, the spectrum of bosonic string contain only massive
particles.

\section{Conclusions}

The three-string and four-string scattering amplitudes were examined for the construction of a field theoretical model 
to reproduce the same scattering amplitudes in the tachyon sector. In string theory cubic couplings of particles 
are completely fixed by expanding the three-string scattering amplitude: Choosing tachyons and gauge particles 
for external string states, the cubic interaction terms between tachyons and gauge particles were obtained. Local quartic interactions between particles can be determined by four-string scattering amplitudes:
Four-string scattering amplitudes contain both four-particle scattering amplitudes generated 
perturbatively by cubic particle interaction and those due to local four-particle interactions. Thus, 
in order to obtain local four-particle interactions, the four-particle scattering amplitudes
generated perturbatively by cubic interactions must be subtracted from the four-particle scattering amplitude deduced from the 
four-string scattering amplitudes. In previous works, procedure was applied to four-gauge-field scattering 
amplitude and obtained the quartic gauge interaction term of non-Abelian gauge theory from the four-string 
scattering amplitude on multiple $Dp$-branes \cite{Lee2016i,Lee2017d}. In this work, the Veneziano amplitude
was decomposed into the four-particle scattering amplitudes by using Feynman diagrams. If the Veneziano amplitude
is expanded in terms of $s$-channel poles, each term may be identified with the corresponding particle scattering amplitude
depicted by a Feynman diagram. To summarize the result, we identify the 
precise form of the quartic tachyon interaction, which may lead the tachyon to condensate and stablizes 
the $D$-brane system. The tachyon fields acquire masses at Plank scale and completely disappear from the 
low energy spectrum. 

It is interesting to note that gauge fields also become massive due to a cubic interaction 
$\phi A_\m A^\m$ after the tachyon condensation. In usual Higgs mechanism, gauge fields become massive 
through the quartic interaction of type $\vert\phi\vert^2 A_\m A^\m$. (The masses of gauge fields 
may also get additional contributions from the quartic interaction, $\vert\phi\vert^2 A_\m A^\m$.) 
It may be more interesting to analyze the tachyons in open string theories on multiple $Dp$-branes and tachyon condensation on more
sophisticated configurations of $D$-brane systems such as $D$-$\bar D$ systems 
\cite{sen1998,KrausL2001,TakayanagiU2001,Hatefi2013,Hatefi2016,Hatefi2017,Hatefi2019,Gutperle2002,Sakai2005} 
by extending this work. Vacuum of string theories on multiple $Dp$-branes may have richer structures than currently known. Certainly, the closed string tachyon must be also on a short list to be explored along the line we discussed here. Some results of further explorations in these directions 
may be presented in papers sequel to this work.

\vskip 1cm

\begin{acknowledgments}
This work was supported by Basic Science Research Program through the National Research Foundation of Korea(NRF) funded by the Ministry of Education (2017R1D1A1A02017805). The author acknowledges the 
the hospitality at APCTP where part of this work was done. 
\end{acknowledgments}


\end{document}